\documentclass[10pt,letterpaper,twocolumn]{article} 

\usepackage{ol2}
\usepackage[draft]{hyperref}
\usepackage{amsmath}

\begin{document}

\twocolumn[ 

\title{Modal dynamics in multimode fibers}


\author{Moti Fridman, Haim Suchowski, Micha Nixon, Asher A. Friesem, and Nir Davidson$^*$}

\address{
Weizmann Institute of Science, Dept. of Physics of Complex Systems,
Rehovot 76100, Israel \\ $^*$Corresponding author:
fefridma@wisemail.weizmann.ac.il }

\begin{abstract}
The dynamics of modes and their states of polarizations in multimode fibers as a function of time, space, and wavelength are experimentally and theoretically investigated. The results reveal that the states of polarizations are displaced in Poincare sphere representation when varying the angular orientations of the polarization at the incident light. Such displacements, which complicates the interpretation of the results, are overcome by resorting to modified Poincare spheres representation. With such modification it should be possible to predict the output modes and their state of polarization when the input mode and state of polarization are known.
\end{abstract}


 ] 

\noindent

%
%
%
%
%
%
%
%
%


\section{Introduction}

Multimode fibers have high coupling efficiencies, can operate over a wide range of wavelengths, and have relatively low susceptibility to degrading nonlinear effects. As a result they are widely used in mid and short communication ranges, high power lasers and amplifiers, and in transporting high optical powers from one location to another~\cite{MMcom1, MMcom2}. Yet, while in single mode fibers the polarization dispersion is low and can be significantly reduced either by using polarization maintaining fibers or by operating at the principal state of polarization~\cite{PMDreview, PMDMeasure, PSP86, Kogelnik, dynamicPSP, dynamicPSP2, dynamicPSP3}, in multimode fibers the mode dispersion is high resulting in significant reduction of performance~\cite{AgrawallBook, Wielandy}. Consequently, it is important to continuously analyze and monitor the dynamics of the modes propagating in multimode fibers.

When analyzing the dynamics of modes in multimode fibers, it is common to separate the dynamics of the transverse modes and the dynamics of the states of polarization~\cite{multi_poincare_Padgett, multi_poincare_agrawal, KahnOL, Kahn2}. Unfortunately, since there is coupling between transverse modes with different states of polarization, such a separation is not natural and results in coupled equations that are difficult to analyze. Here we present a new approach where the complicated dynamics of the modes in multimode fibers are separated and represented by two uncoupled equations of motion that lead to simple geometric plots on modified and uncoupled Poincare spheres. Such plots fully describe the modal dynamics in multimode fibers, and can be conveniently used for monitoring, analyzing and predicting them.

We begin by describing our experimental configuration and present experimental results of the state of polarization (SOP) at the output from multimode fiber as a function of wavelength for different angular orientations of light with linear polarization at the input. The results reveal that the dynamics of the modes and their SOP at the output from multimode fibers are relatively complicated. Then we developed an analytical model for calculating these dynamics, and show how their representation can be simplified to obtain a more convenient and intuitive understanding. Finally, based on the results from the model, we show how to obtain such simplified representation experimentally.

\section{Experimental configuration and results}

\begin{figure}[htb]
\centerline{\includegraphics[width=8.3cm]{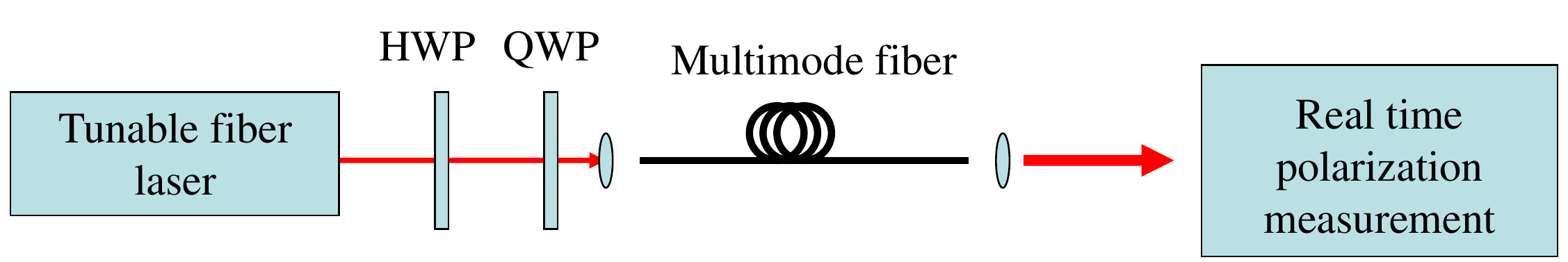}}
\caption{Experimental configuration for measuring the output polarization from a multimode fiber as a function of the input wavelength for different input states of polarization.}\label{system}
\end{figure}

\begin{figure}[htb]
\centerline{\includegraphics[width=5cm]{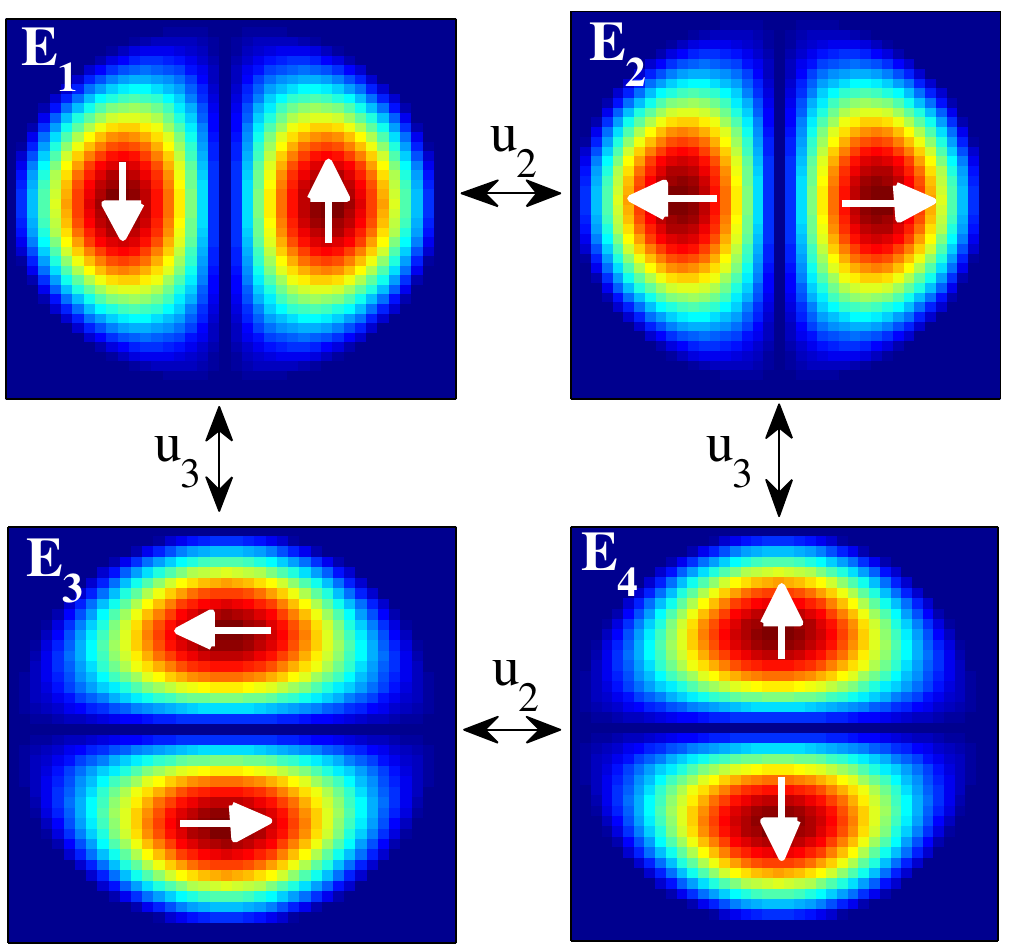}}
\caption{Intensity distributions together with the orientations of the polarization direction for the orthogonal sets of linearly polarized modes $LP_{11}$. The white arrows in the center of the individual intensity distributions denote the polarization directions. The black arrows and the terms $u_2$ and $u_3$ denote the coupling between the different modes.}\label{LP}
\end{figure}

The experimental configuration for measuring the output polarization from a multimode fiber is presented in Fig.~\ref{system}. A linearly polarized light from a tunable fiber laser propagates through a half wave plate and a quarter wave plate in order to obtain light of any desired SOP and wavelength at the input. The light is then launched into a multimode fiber, and the SOP at the output as a function of space, time and wavelength is measured using a real-time space-variant polarization measurement system~\cite{RTSVPM}. In our experiments the multimode fiber length was $10m$, core diameter $20 \mu m$ and numerical aperture 0.07. It mainly supported the $TE_{01}$ and $TM_{01}$ modes which in the weakly guided approximation can be expressed as orthogonal sets of linearly polarized modes. The electric field intensity distributions and polarization directions for these modes are presented in Fig.~\ref{LP}. As shown, there are four distinct modes of $LP_{11}$, each with different intensity distribution or different orientations of the polarization direction. The white arrows at the center of the individual intensity distributions denote the direction of the polarizations in each mode.

We started our experiments by verifying that the coupling into the fiber does not change the mode or the SOP of the input. This was done by determining that the intensity distribution of the mode and the SOP at the output are the same as those at the input after propagating through a short $7cm$ fiber. Then, we launched the same linearly polarized light with a uniform intensity distribution into the 10m long fiber, and measured the SOP at the output while varying the input wavelength from $1062nm$ to $1066nm$. We repeated these measurements, each time after rotating the linearly polarized light at the input by $10^\circ$.

\begin{figure}[htb]
\centerline{\includegraphics[width=8.3cm]{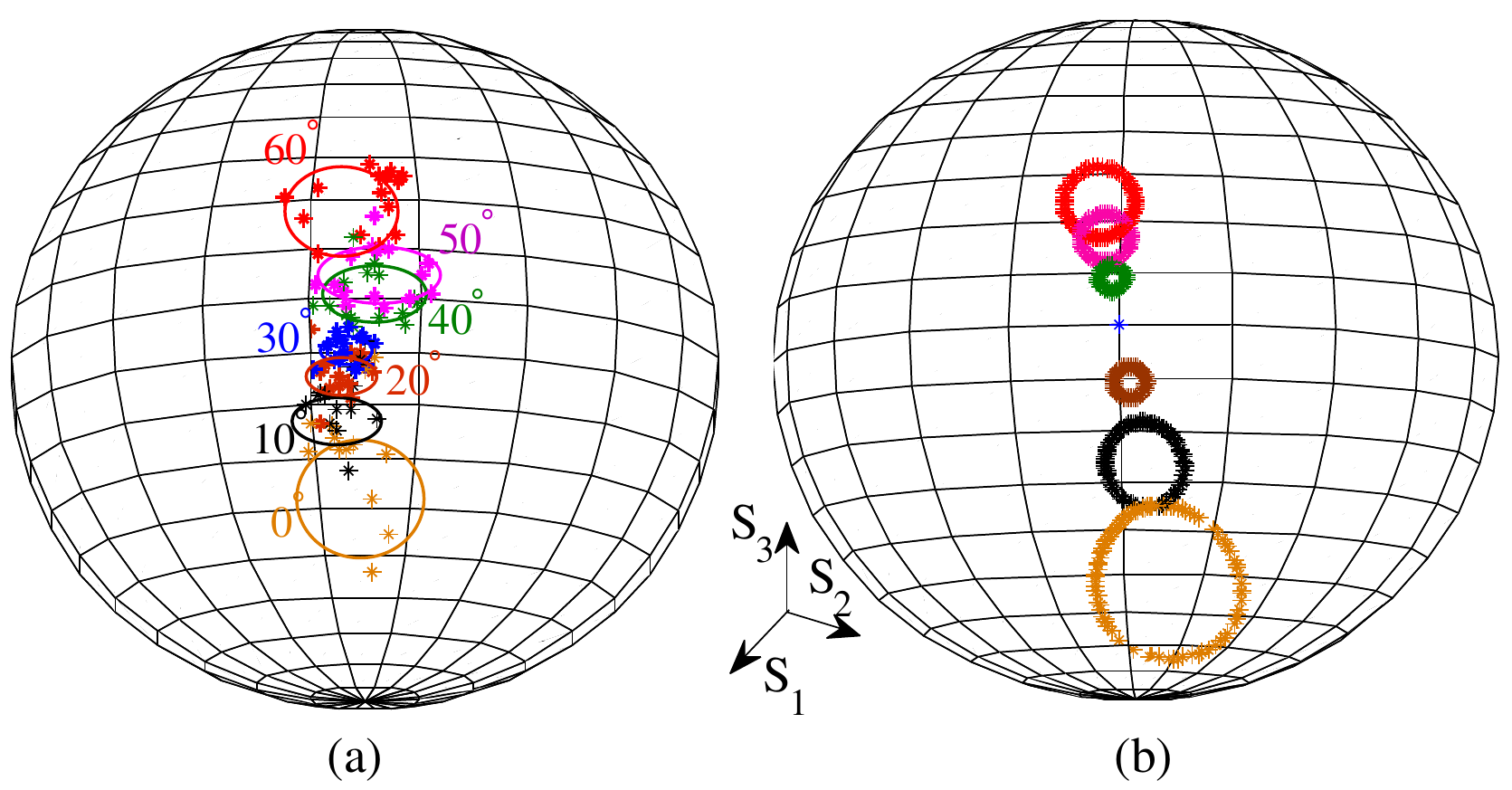}}
\caption{Poincare sphere representation of the SOP at the output as a function of the wavelength for different angular orientations of the linear polarization at the input. (a) Experimental results; (b) calculated results. The input wavelength ranged from $1062nm$ to $1066nm$ for each angular orientation.}\label{ExPass}
\end{figure}

\begin{figure}[htb]
\centerline{\includegraphics[width=8.3cm]{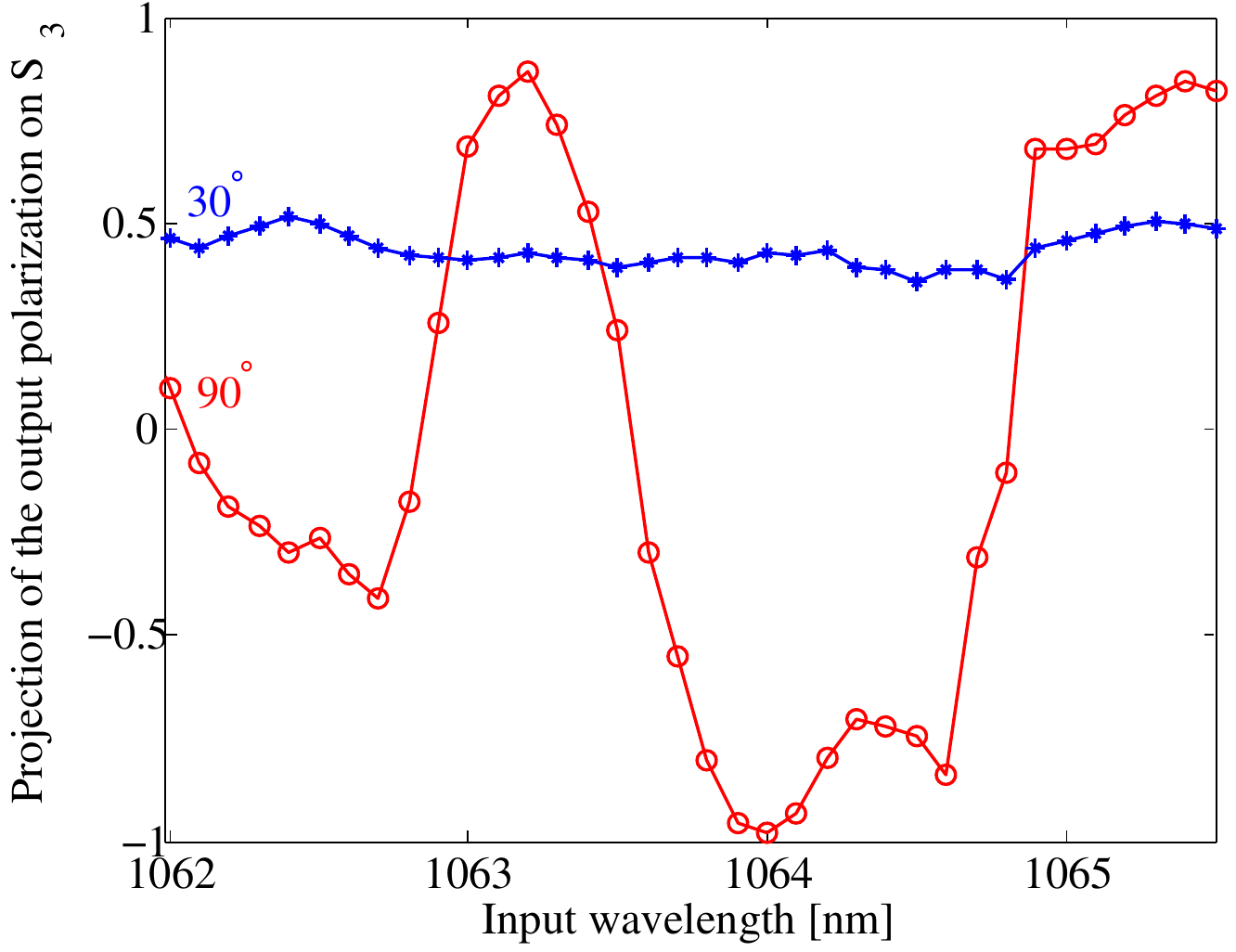}}
\caption{Projection of the output polarization on the $S_1$ axes as a function of the input wavelength for two orientations of the input polarization; asterisks (blue) - $30^\circ$ orientation, close to the PM; circles (red) - $90^\circ$ orientation, far from the PM.}\label{projec}
\end{figure}

The results are presented in Fig.~\ref{ExPass} as a Poincare sphere representation of the SOP at the output for different angular orientations of the linear polarization at the input. As evident, the circle of smallest size is obtain when the linearly polarized input beam is oriented at $30^{\circ}$, indicating where the SOP of the output is nearly invariant to the input wavelengths. Namely, at an angular orientation of about $30^\circ$ we have a principal mode (PM). As the deviation from $30^\circ$ increases, the circles become larger indicating that more spectral dispersion occurs. Unlike increasing concentric circles around the principal state of polarization location in single mode fiber, here we obtained increasing circles whose centers continuously deviate from the PM location as the difference between the input and the PM increases. We also present the projection of the output polarization on the $S_1$ axes as a function of the input wavelength for two input polarization in Fig.~\ref{projec}. As evident, when the input polarization is close to the PM the output polarization is less sensitive to wavelength variations.

\section{Model and calculated results}

We developed a model for calculating the dynamics of the SOP at the output of multimode fibers, in order to gain a better physical intuition about the evolution of modes in a multimode fiber and to understand what causes the deviations from the PM location. We define $E_{in}$ and $E_{out}$ as the input and output modes composed from a superposition of the four fields $E_{in,out}=[E_1,E_2,E_3,E_4]^T_{in,out}$ shown in Fig.~\ref{LP}. The relation between them is
\begin{equation}
E_{out}=e^{i \beta} \mathbf{U} E_{in},
\label{EqDynamic}
\end{equation}
where $\beta$ is real and $\mathbf{U}$ is a unitary matrix. To explicitly define $\mathbf{U}$, we assume that birefringence and twists in the fiber are the main sources for coupling between the modes. Birefringence in the fiber leads to coupling of modes with the same intensity distributions but with orthogonal SOP. Accordingly, the coupling between the fields $E_1$ and $E_2$ can be described by a 2 by 2 matrix, as
\begin{equation}
\mathbf{U}^{(2)}(\omega)=\left[ \begin{array}{cc}
  u_1(\omega)    & u_2(\omega)   \\
  -u_2^*(\omega) & u_1^*(\omega) \\
\end{array} \right] ,
\end{equation}
where $u_1$ denote changes in the phase of each mode while propagating in the fiber and $u_2$ the coupling strength between the modes. This matrix is the same as that for coupling orthogonal polarizations in single mode fibers~\cite{PSP86}. Similarly, since $E_3$ and $E_4$ also have the same intensity distributions and assuming that birefringence is essentially constant across a relatively small fiber core~\cite{Ulrich79, tweestedWG}, we can describe the coupling between $E_3$ and $E_4$ with the same unitary matrix $U^{(2)}(\omega)$. Twists in the fiber rotate the modes and leads to coupling between $E_1$ and $E_3$ and between $E_2$ and $E_4$ with coupling strength denoted as $u_3$. We also assume that coupling between $E_1$ and $E_4$ and between $E_2$ and $E_3$ are negligible. Taking birefringence and twists into account, the overall coupling matrix between all four fields is then $\mathbf{U}(\omega)=\exp (\mathbf{H}) $ where $\mathbf{H}$ is skew-Hermitian matrix as
\begin{equation}
\mathbf{H}(\omega)=\left[ \begin{array}{cccc}
  u_1(\omega)    & u_2(\omega)   & u_3(\omega)  & 0\\
  -u_2^*(\omega) & u_1^*(\omega) & 0 & u_3(\omega) \\
  -u_3^*(\omega) & 0 & u_1(\omega)    & u_2(\omega)\\
  0 & -u_3^*(\omega) & -u_2^*(\omega) & u_1^*(\omega) \\
\end{array} \right].
\end{equation}
Following the derivation of Poole and Wagner~\cite{PSP86} and extending it to our multimode fibers~\cite{KahnOL, Kahn2}, leads to four principal modes (PMs) where each is insensitive to variations in the input wavelength up to the first order. The four PMs, denoted as $\epsilon_{n}$ where $n=1,2,3$ and 4, are the solutions of
\begin{equation}
\left[ \mathbf{U}'-i  k_{n} \mathbf{U} \right] \epsilon_{n}=\left[ \mathbf{H}-i k_{n} \mathbf{I} \right] \mathbf{U} \epsilon_{n}=0,
\label{EqMatEv}
\end{equation}
where the prime denotes differentiation with respect to frequency, and the four $k_n$ are
\begin{equation}
k_{n}=\pm i \sqrt{|u_1|^2+|u_2|^2+|u_3|^2 \pm 2 |u_3| \sqrt{|u_1|^2+|u_2|^2}}.
\end{equation}
These four PMs will propagate in the multimode fiber with minimal dispersion.

To analyze the dynamics of modes other than PMs as a function of the input wavelength, we differentiated Eq.~(\ref{EqDynamic}) with respect to frequency, and used the definition of $\mathbf{U}$ to obtain
\begin{equation}
\frac{d E_{out}}{d \omega}=\mathbf{H} E_{out}.
\label{EqEv}
\end{equation}
Using Pauli matrices, $\sigma_x$, $\sigma_y$ and $\sigma_z$, we can decompose $\mathbf{H}$ to
\begin{equation}
\mathbf{H} =  \left( i  b_1 \sigma_z+i a_2 \sigma_y + i b_2 \sigma_x \right) \otimes I + I \otimes \left(i b_3 \sigma_x + i a_3 \sigma_y \right),\label{EqPauli}
\end{equation}
where $a_i$ denote the real part of $u_i$ and $b_i$ the imaginary part of $u_i$. Equation (\ref{EqPauli}) defines rotation in four dimensions, one part for left rotation and the other for right rotation~\cite{HaimThesis, DuVal}. Now, in order to find analytic solution to Eq.~(\ref{EqEv}) we redefine the vectors $E_{in}$ and $E_{out}$ to a 2 by 2 matrices,
\begin{equation}
\tilde{E}_{in,out}= \left[ \begin{array}{cc}
  E_1 & E_2 \\
  E_3 & E_4 \\
\end{array} \right]_{in,out},\label{EqE2x2}
\end{equation}
and substituting Eqs.~(\ref{EqPauli}) and~(\ref{EqE2x2}) into Eq.~(\ref{EqMatEv}) leads to the solution of $\tilde{E}_{out}$ as a function of frequency as
\begin{equation}
\tilde{E}_{out}(\omega)=\mathbf{\Omega}_L(\omega) \tilde{E}_{in}
\mathbf{\Omega}_R(\omega), \label{anaSol}
\end{equation}
where
\begin{equation}
\mathbf{\Omega}_L(\omega)= \left[ \begin{array}{cc}
  \cos(|\tau_b| \omega) & \frac{\tilde{u}_3}{|\tau_b|}\sin(|\tau_b| \omega)  \\
  -\frac{\tilde{u}^*_3}{|\tau_b|}\sin(|\tau_b| \omega)  & \cos(|\tau_b| \omega) \\
\end{array} \right] \nonumber
\end{equation}
and
\begin{eqnarray}
 \mathbf{\Omega}_R(\omega) = \ \ \ \ \ \ \ \ \ \ \ \ \ \ \ \ \ \ \ \ \ \ \ \ \ \ \ \ \ \ \ \ \ \ \ \ \ \ \ \ \ \ \ \ \ \ \ \ \ \ \ \ \ \nonumber \\
 \left[\begin{array}{cc}
  \cos(|\tau_a| \omega) + \frac{\tilde{u}_1}{|\tau_a|} \sin(|\tau_a| \omega) & \frac{-\tilde{u}^*_2}{|\tau_a|}\sin(|\tau_a| \omega)  \\
  \frac{\tilde{u}_2}{|\tau_a|}\sin(|\tau_a| \omega) & \cos(|\tau_a| \omega) - \frac{\tilde{u}_1}{|\tau_a|} \sin(|\tau_a| \omega)  \\
\end{array} \right] \nonumber
\end{eqnarray}
with $\tau_a \equiv (b_1,b_2,a_2)$ and $\tau_b \equiv (0,a_3,b_3)$.

Equation (\ref{anaSol}) provides an analytic solution of the output mode distribution and its SOP as a function of the wavelength when the input SOP is known. It involves a multiplication of two matrices $\mathbf{\Omega}_R$ and $\mathbf{\Omega}_L$. The matrix $\mathbf{\Omega}_R$ is composed from $u_1$ and $u_2$ so it represents the coupling between each pair of modes with the same intensity distributions but with different SOP whereas the matrix $\mathbf{\Omega}_L$ is composed only from $u_3$ so it represents the coupling between each pair of modes with different intensity distributions and different SOP. Using Eq.~(\ref{anaSol}), we calculated the evolution of the output SOP as a function of the input wavelength for different angular orientations of a linearly polarized light at the input. The results are presented in Fig.~\ref{ExPass}, with the coupling parameters $u_1$, $u_2$ and $u_3$, as fitting parameters. As evident, there is a qualitative agreement between the experimental and calculated results, indicating that our assumptions in developing the model are justified. We attribute the difference between the calculated results and the measured results to small deviations in the polarization measurement and to inaccurate coupling parameters.

\section{Modified Poincare spheres representation}

In order to simplify the Poincare sphere representation of the SOP dynamics at the output, we resorted to modified Stokes parameters $S_a \equiv (S_a^{(1)},S_a^{(2)},S_a^{(3)})$ and $S_b \equiv (S_b^{(1)},S_b^{(2)},S_b^{(3)})$ where
\begin{equation}
\begin{array}{cr}
S_a^{(1)}=|E_1|^2-|E_2|^2+|E_3|^2-|E_4|^2 \\
S_a^{(2)}=2|E_1||E_2|\cos \left( \varphi_{12} \right) + 2|E_3||E_4|\cos \left( \varphi_{34}\right) \\
S_a^{(3)}=2|E_1||E_2|\sin \left( \varphi_{12} \right) + 2|E_3||E_4|\sin \left( \varphi_{34}\right)
\end{array} \label{EqA}
\end{equation}
and
\begin{equation}
\begin{array}{cl}
S_b^{(1)}=|E_1|^2+|E_2|^2-|E_3|^2-|E_4|^2 \\
S_b^{(2)}=2|E_1||E_3|\cos \left( \varphi_{13} \right) + 2|E_2||E_4|\cos \left( \varphi_{24}\right) \\
S_b^{(3)}=2|E_1||E_3|\sin \left( \varphi_{13} \right) + 2|E_2||E_4|\sin \left( \varphi_{24}\right)
\end{array}
\end{equation}
where $\varphi_{ij}$ denote the phase difference between $E_i$ and $E_j$. Now, Eq.~(\ref{EqEv}) can be separated into two uncoupled torque equations, as
\begin{equation}
\begin{array}{cr}
\dot{S}_a=\tau_a \times S_a \\
\dot{S}_b=\tau_b \times S_b.
\end{array}
\end{equation}
Note, the torque vectors depend only on the parameters of the fiber and not on the input SOP.

Using the two sets of modified Stokes parameters we recalculated the evolution of the output SOP as a function of the input wavelength for different angular orientations of a linearly polarized light at the input. The results are presented in Fig.~\ref{numericscor}. As evident, the use of the modified Stokes parameters and the separation of Eq.~(\ref{EqEv}) into two uncoupled equations, resulted in two modified Poincare spheres where the SOPs in each sphere form concentric, rather than displaced, circles around the PM when varying the input wavelength. The modified Poincare spheres representations involve natural modes of a fiber. Specifically, $S_a$ represents the modes $TE_{01}$ and $TM_{01}$, and $S_b$ represents the modes $HE_{21}$ and $HE_{-21}$. These results indicate that when evaluating the propagation of light in multimode fiber it is preferable and more convenient to resort to our uncoupled modified Poincare spheres representation rather than the usual Poincare spheres representation~\cite{multi_poincare_Padgett, multi_poincare_agrawal}.

\begin{figure}[htb]
\centerline{\includegraphics[width=8cm]{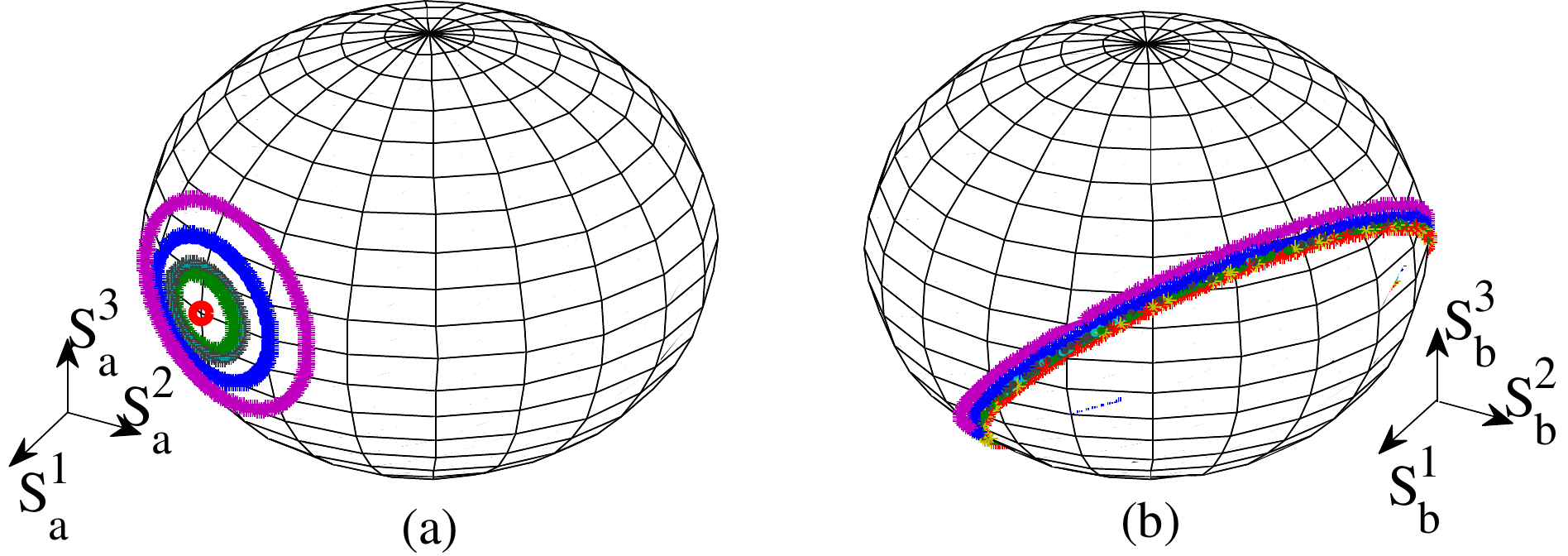}}
\caption{Calculated modified Poincare sphere representation of the SOP at the output as a function of the wavelength for different angular orientations of linearly polarized light at the input. (a) Using the modified Stokes parameters $S_a$; (b) using the modified Stokes parameters $S_b$.}
\end{figure}\label{numericscor}

Based on the calculated results from our model we reevaluated our experimental results in order to obtain more convenient and more intuitive Poincare sphere representations of the SOP dynamics at the output from a multimode fiber. Specifically, we first multiplied the measured light distribution with those of the transverse modes shown in Fig.~\ref{LP}. Then we determine the SOP at each point across the beam which leads to determining the contribution of the fields $E_1$, $E_2$, $E_3$, and $E_4$. Next, we used Eq.~(\ref{EqA}) to determine the modified Stokes parameters $S_a$ to obtain the desired modified Poincare sphere representation. For simplicity we present only the average SOP (center of circles) for each angular orientation of the linearly polarized input beam. The results are presented in Fig.~\ref{numericscor}. Figure~\ref{numericscor}(a) shows the average SOP for different input polarization orientations in the original Poincare sphere representation, same as Fig~\ref{ExPass}. Here, the average SOP is displaced as we vary the input polarization orientation. Figure~\ref{numericscor}(b) shows the corresponding results on the modified Poincare sphere representation. As evident, here the average SOP are about the same for the different input polarization orientations, as expected.

\begin{figure}[htb]
\centerline{\includegraphics[width=8cm]{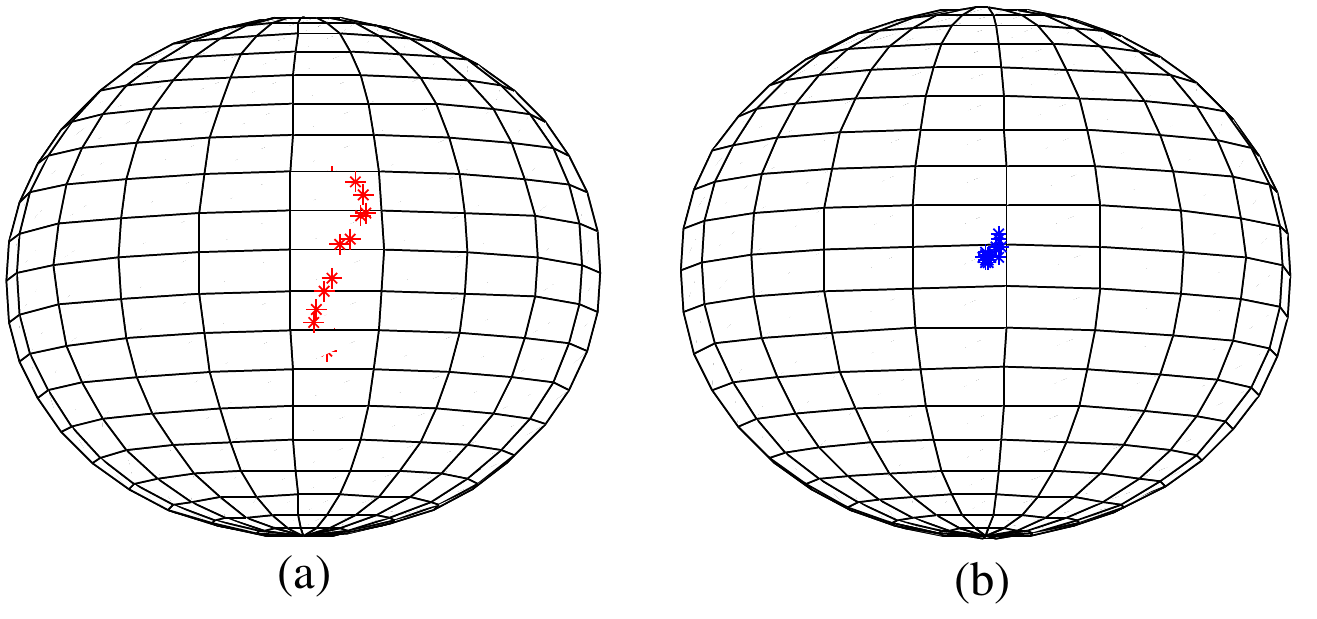}}
\caption{The average polarization when varying the input wavelength for different input polarizations on (a) the original Poincare sphere and on (b) the modified Poincare sphere.}\label{numericscor}
\end{figure}

\section{Concluding remarks}

To conclude, we presented experimental and calculated results about the dynamics of modes and their SOP in multimode fibers. We show how the representation of these dynamics can be simplified so as to obtain a direct measure of the SOP at the output when knowing the input SOP. The results of this measure can then be used for determining the SOP for an arbitrary input SOP.

\section*{Acknowledgements}

This research was supported by the Israeli Ministry of Science and Technology and by the USA-Israel Binational Science Foundation.


\end{document}